\documentclass[a4paper,11pt]{article}
\usepackage[utf8]{inputenc}
\pdfoutput=1
\usepackage{jheppub,float}
\usepackage[normalem]{ulem}
\usepackage{xcolor}
\usepackage{amsmath}
\usepackage{framed}
\colorlet{shadecolor}{lightgray}
\usepackage{hyperref}
\usepackage{parskip}
\usepackage{amsmath, amssymb, amsthm, graphicx, epsfig, fancyhdr,epsfig}
\usepackage{float}
\usepackage{xcolor}

\newcommand{\be}{\begin{equation}}
\newcommand{\ee}{\end{equation}}
\newcommand{\bea}{\begin{eqnarray}}
\newcommand{\eea}{\end{eqnarray}}
\newcommand{\cphi}{\varphi}

\newcommand{\cH}{\mathcal{H}}

\newcommand{\non}{\nonumber}
\newcommand{\cf}{\mathfrak{f}}

\title{\boldmath Inflation and primordial gravitational waves in scale-invariant quadratic gravity with Higgs}

\author[1]{Anish Ghoshal,}
\author[2,3]{Debangshu Mukherjee,}
\author[4,5]{Massimiliano Rinaldi.}

\affiliation[1]{Institute of Theoretical Physics, Faculty of Physics, University of Warsaw, ul. Pasteura 5, 02-093 Warsaw, Poland}
\affiliation[2]{School of Physics, Indian Institute of Science Education \& Research Thiruvananthapuram, Vithura 695551, Kerala, India}
\affiliation[3]{Department of Physics, Indian Institute of Technology Kanpur, Kanpur 208016, India}
\affiliation[4]{Department of Physics, University of Trento, Via Sommarive 14, 38123 Trento, Italy}
\affiliation[5]{TIFPA-INFN, Via Sommarive 14, 38123 Trento, Italy}
\vspace*{0.3cm}
\emailAdd{anish.ghoshal@fuw.edu.pl}
\emailAdd{debangshu@iitk.ac.in}
\emailAdd{massimiliano.rinaldi@unitn.it}

\abstract{In scale-invariant models of fundamental physics all mass scales are generated via spontaneous symmetry breaking. In this work, we study inflation in scale-invariant quadratic gravity, in which the Planck mass is generated classically by a scalar field, which evolves from an unstable fixed point to a stable one thus breaking  scale-invariance. We investigate the  dynamics by means of dynamical system standard techniques. By computing the spectral indices and comparing them with data, we put some constraints on the three dimensionless parameters of the theory. We show that certain regions of the parameter space will be within the range of future CMB missions like CMB-S4, LiteBIRD and STPol. The second half of the paper is dedicated to the analysis of inflationary first-order tensor perturbations and the calculation of the power spectrum of the gravitational waves. We comment on our results and compare them with the ones of mixed Starobinsky-Higgs inflation.}

\begin{document} 

\maketitle
\flushbottom

\section{Introduction}
Understanding a dynamical origin of energy scales (like the Planck scale or the electroweak scale) has been a quest for theoretical research in gravity and field theory. Scale-invariant theories present several very interesting aspects in this perspective as they do not contain any fundamental scale in the action at the classical level but dynamically leads to scale-genesis via quantum corrections \cite{Cooper:1981byv,RevModPhys.54.729,PhysRevD.7.1888,Salvio:2014soa,Einhorn:2014gfa,Einhorn:2015lzy,Einhorn:2016mws,Edery:2015wha,Ferreira:2018qss,Ferreira:2019zzx,Salvio:2020axm,Ferreira:2021ctx}. Some of the salient features that are manifest from such investigations over the past four decades lead to naturally flat inflationary potentials  \cite{Khoze:2013uia,Kannike:2014mia,Rinaldi:2014gha,Salvio:2014soa,Kannike:2015apa,Barrie:2016rnv,Tambalo:2016eqr} and provide mechanisms of particle production or dark matter candidates  \cite{Hambye:2013dgv,Karam:2015jta,Kannike:2015apa,Kannike:2016bny,Karam:2016rsz,Plascencia:2015xwa,Khoze:2016zfi}, possess a motivating framework to address the gauge hierarchy problem, \cite{Bardeen,Foot:2007iy,Alexander-Nunneley:2010tyr,Englert:2013gz,Hambye:2013dgv,Farzinnia:2013pga,Altmannshofer:2014vra,Holthausen:2013ota,Salvio:2014soa,Einhorn:2014gfa,Kannike:2015apa,Farzinnia:2015fka,Kannike:2016bny,Ghoshal:2017egr,Ghoshal:2018gpq,Frasca:2020ojd,Ghoshal:2020lfd,Frasca:2020jbe,Barman:2021lot,Frasca:2022kfy,Ghoshal:2022hyc,Barman:2022njh} and also leads to strong first-order phase transitions in early universe. In turn, these can produce large amplitude gravitational wave (GW) signals \footnote{For inflationary sources of GWs in modified gravity theories, see Refs. \cite{Bernal:2020ywq,Odintsov:2022cbm,Odintsov:2021kup} and for dark matter\cite{Ghoshal:2021ief}.}, mainly due to the dominant nature of thermal corrections in the absence of tree-level mass terms ~\cite{Jaeckel:2016jlh, Marzola:2017jzl, Iso:2017uuu, Baldes:2018emh, Prokopec:2018tnq, Brdar:2018num, Marzo:2018nov, Ghoshal:2020vud}.
Black holes in quadratic gravity were studied concerning their stability in \cite{Dioguardi:2020nxr}, the thermodynamics in \cite{Cognola:2015uva,Cognola:2015wqa}, and as dark matter candidates in \cite{Aydemir:2020xfd}.  Finally the importance of such theories having no-scales also lies in the fact that the only dimension-4 operators are allowed in the classical Lagrangian and this means that there is a strong constraint on the allowed free parameters in the action, which in context to the gravitational piece of the Lagrangian says that it is only quadratic in the curvature tensors and not beyond. The most general gravitational Lagrangian can be shown to be the sum of the Ricci scalar squared $R^2$ and the Weyl-squared terms. 

Thus, if Nature does not allow us to have fundamental scales at the classical level and all observed energy scales are in fact dynamically generated (like via quantum corrections) then it gives us a theory of gravity containing only terms quadratic in the curvature along with all possible scale-invariant couplings to the matter sector. Now, since all scales will be generated at 1-loop level, this puts constraints on some of the parameter space and makes predictions in the early universe, as in inflation.

Particularly, in the context of inflationary cosmology, scale-invariant models of gravity have been able to predict values very consistent with observations for the spectral index of scalar perturbations $n_s$\footnote{In fact in the original Starobinsky model, the fact the $n_s$ departs slightly from 1 is attributed to the Einstein-Hilbert dimensionful term.}. Moreove, scale-invariance symmetry must be broken dynamically, usually by quantum corrections (for example see \cite{zee1979broken,starobinsky_new_1980, gorbunov2014scale,turchetti1981gravitation,Salvio:2014soa,Rinaldi:2014gha,Rinaldi:2015yoa,Barrie:2016rnv})

In the model presented in  Ref \cite{Rinaldi:2015uvu}, scale-invariance is promoted to a global symmetry of the Lagrangian and scale-invariant operators are built with combinations of $R$, $R^{2}$ and a scalar field $\phi$ with a canonical kinetic term and a quartic potential. When implemented with a flat Robertson-Walker metric, the equations of motion show that the corresponding dynamical system has only two fixed points: one is unstable and corresponds to a quasi de-Sitter space-time with a vanishing scalar field while the other, manifestly stable, corresponds to damped oscillations of the Hubble parameter and of the scalar field around non-vanishing fixed values, which depends on the three free parameters of the theory. The most interesting aspect is that the path from the unstable to the stable point corresponds to an arbitrarily long expanding quasi-de Sitter phase followed by damped oscillations. Thus the interpretation of this model as inflation and reheating is quite natural. The (asymptotic) equilibrium value of $\phi$ is associated with a fundamental energy scale, which emerges dynamically (just due to transition from one fixed point to another fixed point in the entire system). Thus the global scaling symmetry breaking spontaneously generates a mass scale in the system, without the need of quantum corrections. This can be the Planck mass that we observe in gravity. Later on the scalar field oscillates and produces excitations of the Standard Model fields leading to successful reheating of the visible Universe that we observe today (see \cite{Rinaldi:2015uvu},  \cite{Vicentini:2019etr}, and \cite{Tambalo:2016eqr} for details). 

In this work we review the model described in \cite{Rinaldi:2015uvu} by considering the three parameters completely free. In fact, in the original model one parameter was a priori fixed in terms of the other two to guarantee certain cancellations in the action at late times. We study in great detail the inflationary phase in the Einstein frame by computing the spectral indices. We also study the generation of primordial gravitational waves and compare our results with similar models \cite{Salvio:2017xul, Myung:2014cra}.

The paper is organized as follows: in Sec.\ \ref{action} we present the classical scale-invariant action underlying the inflationary model we want to study. In Sec.\ \ref{fp} we study the equations of motion by using a dynamical system approach. We find the fixed points and analyse their stability. In Sec.\ \ref{fieldsred} we show that, in the Einstein frame, a suitable field redefinition yields a potential that depends on one field only. In Sec.\ \ref{infl} we study the inflationary evolution, calculate the spectral indices and show how the parameters of the theory are constrained by observations. In Sec.\ \ref{tpert} we compute the spectrum of tensor perturbations for this model and we conclude in Sec.\ \ref{concl} with a summary and a brief discussion of our results.

\section{Scale-invariant model of quadratic gravity}\label{action}

Following  \cite{Rinaldi:2015uvu}, we consider the scale-invariant action
\begin{equation}
\label{eq:fullquadraticaction}
    S=\int d^4x \sqrt{-g}\left[\frac{\alpha}{36}R^2+\frac{\xi}{6}\phi^2R-\frac{1}{2}\partial_{\mu}\phi\partial^{\mu}\phi-\frac{\lambda}{4}\phi^4 \right]\ .
\end{equation}

Unlike \cite{Rinaldi:2015uvu}, however, here we will keep the parameter $\alpha$ arbitrary. By assuming the flat Friedmann-Robertson metric $ds^2=-dt^2+a(t)^2d\vec x^2$ one finds that there are two equations of motion, one for $\phi$ and the other for $H=\dot a/a$,  that contains derivatives up to second order only:
 \bea\label{eomt}
&&\ddot\phi+3H\dot \phi-2\xi\phi\dot H-\phi(4\xi H^{2}-\lambda\phi^{2})=0\,,\\\non
&& \alpha\left(2H\ddot H  -\dot H^{2}+6H^{2}\dot H  \right)-{1\over 2}\dot\phi^{2}+2\xi\phi \dot \phi H+{\phi^{2}\over 4}(4\xi H^{2}-\lambda \phi^{2})=0\,,
\eea
The addition of a term like $\beta R_{\mu\nu}R^{\mu\nu}$ to the action \eqref{eq:fullquadraticaction}  would simply yield the shift $\alpha\rightarrow \alpha+12\beta$ and we do not consider this term here.  Of course, at the perturbative level, the effects of the two parameters $\alpha$ and $\beta$ would probably decouple. However, we choose to set $\beta=0$ to keep the parameter space as small as possible and, at the same time, to capture the essential features of scale-invariant inflation. In addition, a quadratic term in the Ricci tensor would make rather involved the analysis of this model in the Einstein frame, which is one our main goals.
In addition, we do not consider the quadratic Riemann term since it can be factored out by using the Gauss-Bonnet combination in four dimensions.

The action \eqref{eq:fullquadraticaction} is globally scale-invariant since its form does not change under the transformations
\bea\label{eq:globscaleinv}
\bar g_{\mu\nu}(x)=g_{\mu\nu}(\ell x)\,,\quad \bar\phi(x)=\ell\phi(\ell x)\,,
\eea
where $\ell$ is an arbitrary positive dimensionless constant. In contrast, the action is not invariant under local Weyl (or conformal) transformations. Global scale invariance becomes manifest also at the level of the equations of motion \eqref{eomt}, since they are left unchanged in form by the transformations
\bea
\bar\phi(t)=\ell\phi(\ell t)\,,\quad \bar a(t)=a(\ell t)\,,\quad \bar H(t)={d\ln \bar a(t)\over dt}=\ell H(\ell t)\,.
\eea

We now write the action in the Einstein frame. First, we introduce an auxiliary field $\cphi$ along with an auxiliary variable $\chi$ defined as
\begin{equation}
    \chi =\frac{\alpha \cphi}{18}+\frac{\xi \phi^2}{6}\ ,
\end{equation}
which leads to the following form of Lagrangian density
\begin{equation}
\label{eq:auxiliaryL}
    \frac{\mathcal{L}}{\sqrt{-g}}=\chi R-\frac{\alpha \cphi^2}{36}-\frac{1}{2}\partial_{\mu}\phi\partial^{\mu}\phi-\frac{\lambda}{4}\phi^4\ .
\end{equation}
The equation of motion for the auxiliary field $\cphi$ is simply $\cphi =R$ thus ensuring \eqref{eq:auxiliaryL} indeed reduces to \eqref{eq:fullquadraticaction} on-shell. Operationally, one can think that the dynamics described by the action \eqref{eq:fullquadraticaction} near its saddle can be alternatively well-approximated by the saddle of \eqref{eq:auxiliaryL}. To put it differently, on-shell configurations of the auxiliary scalar $\cphi$ boils down to \eqref{eq:fullquadraticaction}.  In terms of the field variable $\chi$, the action \eqref{eq:auxiliaryL} can be recast in the form
\begin{equation}
    S=\int d^4x\ \sqrt{-g}\left[\chi R-\frac{1}{2}(\partial \phi)^2+\frac{3 \xi}{\alpha}\phi^2 \chi -\left(\frac{\lambda}{4}+\frac{\xi^2}{4\alpha}\right)\phi^4-\frac{9}{\alpha}\chi^2 \right]\,.
\end{equation}
To write the above action in the canonical Einstein-Hilbert form we perform the Weyl transformation 
\begin{equation}
    \tilde{g}_{\mu \nu}=e^{2\omega(x)}g_{\mu \nu}\ .
\end{equation}
Finally, by identifying, 
\begin{equation}
    \omega =\frac{1}{2}\log \frac{2\chi}{M^2}\,,
\end{equation}
where $M$ is an arbitrary parameter having dimension of mass, we find the action in Einstein frame
\begin{equation}
\begin{aligned}
    S_E&=\int d^4x\ \sqrt{-\tilde{g}}\left[\frac{M^2}{2}\tilde{R}-3M^2 \tilde{g}^{\mu \nu}\partial_{\mu}\omega \partial_{\nu}\omega-\frac{1}{2}e^{-2\omega}\tilde{g}^{\mu \nu}\partial_{\mu}\phi\partial_{\nu}\phi+\frac{3M^2\xi}{2\alpha}e^{-2\omega}\phi^2 \right.\\
    &\hspace*{8cm}\left. -\left(\frac{\lambda}{4}+\frac{\xi^2}{4\alpha} \right)e^{-4\omega}\phi^4-\frac{9M^4}{4\alpha} \right]\,.
\end{aligned}
\end{equation}
The kinetic term for $\omega$ can be written in the  canonical form by defining $\tilde{\omega}=\sqrt{6}M\omega$, thus the action reads
\begin{equation}
\label{eq:einsteinframe-action}
    S_E= \int d^4x\ \sqrt{-\tilde{g}}\left[\frac{M^2}{2}\tilde{R}-\frac{1}{2}\tilde{g}^{\mu \nu}\partial_{\mu}\tilde{\omega}\partial_{\nu}\tilde{\omega}-\frac{1}{2}e^{-\sqrt{\frac{2}{3}}\frac{\tilde{\omega}}{M}}\tilde{g}^{\mu \nu}\partial_{\mu}\phi \partial_{\nu}\phi+V(\tilde{\omega},\phi)-\frac{9M^4}{4\alpha} \right]\,,
\end{equation}
where the potential
\begin{equation}
\label{eq:potential}
    V(\tilde{\omega},\phi)=\frac{3M^2\xi}{2\alpha}e^{-\sqrt{\frac{2}{3}}\frac{\tilde{\omega}}{M}}\phi^2-\left(\frac{\lambda}{4}+\frac{\xi^2}{4\alpha} \right)e^{-2\sqrt{\frac{2}{3}}\frac{\tilde{\omega}}{M}}\phi^4 \ ,
\end{equation}
involve the two scalars $\phi$ and $\tilde{\omega}$ coupled to gravity while the constant term appearing in \eqref{eq:einsteinframe-action} can be thought of as a cosmological constant. 

\section{Fixed-point analysis of scale-invariant model}\label{fp}

We now perform a dynamical system analysis of \eqref{eq:einsteinframe-action} to study the fixed points. For computational convenience, we introduce the field $\cf$, defined as,
\bea
\cf=M\,e^{-{{\bar\omega}\over \sqrt{6}M}}\ .
\eea
The Lagrangian density given in \eqref{eq:einsteinframe-action} can then be written as,
\bea
\label{eq:finalL}
\mathcal{L}=\sqrt{g}\left[ {M^{2}\over 2}R-{3M^{2}\over \cf^{2}}(\partial \cf)^{2}-{\cf^{2}\over 2M^{2}}(\partial\phi)^{2}+{3\xi \phi^{2}\cf^{2}\over 2\alpha} -{\Omega\phi^{4}\cf^{4}\over 4\alpha  M^{4}} -{9M^{4}\over 4\alpha}\right]\ ,
\eea
where
\bea
\Omega=\alpha\lambda+\xi^{2}\,.
\eea
Note that in the above, we have dropped the `bar' from $R$ and $g_{\mu \nu}$ since it is understood that we are working with the action in Einstein frame.

By imposing the usual FLRW metric
\begin{equation}
\label{eq:FLRWmetric}
    ds^2=-dt^2+a(t)^2(dx^2+dy^2+dz^2)\,,
\end{equation}
we find the Friedmann equations 
\bea
H^{2}&=& {\dot \cf^{2}\over \cf^{2}}+{\cf^{2}\dot\phi^{2}\over 6M^{4}}+{3M^{2}\over 4\alpha}-{\xi \cf^{2}\phi^{2}\over 2\alpha M^{2}}+{\Omega\cf^{4}\phi^{4}\over 12\alpha M^{6}}\ ,\\\non
\label{eq:friedmanneq}
\dot H&=&-3{\dot \cf^{2}\over \cf^{2}}-{\cf^{2}\dot\phi^{2}\over 2M^{4}}\ ,
\eea
where $H=\frac{\dot{a}}{a}$ is the Hubble parameter, while the Klein-Gordon equations for $\phi$ and $\cf$ take the form
\bea
\ddot\phi+3H\dot\phi+2{\dot\phi\dot \cf\over \cf}+{\Omega\cf^{2}\phi^{3}\over \alpha M^{2}}-{3\xi M^{2}\phi\over \alpha}&=&0\\\non
\label{eq:kleingordoneq}
{\ddot \cf\over \cf}+{\dot \cf\over \cf}\left(3H-{\dot \cf\over \cf}\right)-{\dot\phi^{2}\cf^{2}\over 6M^{4}}-{\xi \cf^{2}\phi^{2}\over 2\alpha M^{2}}+{\Omega\cf^{4}\phi^{4}\over 6 \alpha M^{6}}&=&0\ .
\eea
Here, the dots denote derivative with respect  $t$. To study the fixed points and their stability, we define the new variables
\bea
x_{1}=\cf,\quad x_{2}=\dot \cf, \quad y_{1}=\phi, \quad y_{2}=\dot\phi, \quad H=z\,. 
\eea
The first Friedmann equation reads
\bea\label{aux}
z^{2}={x_{2}^{2}\over x_{1}^{2}}+{x_{1}^{2}y_{2}^{2}\over 6M^{4}}-{\xi x_{1}^{2}y_{1}^{2}\over 2\alpha M^{2}}+{\Omega x_{1}^{4}y_{1}^{4}\over 12\alpha M^{6}}+{3M^{2}\over 4\alpha}\,,
\eea
and we consider it as a constraint. The second one along with the two Klein-Gordon equations and two auxiliary conditions leads to a closed system of non-linear first order equations given by
\bea
\label{eq:fpe1}
\dot z&=&-{3x_2^2\over x_1^2}-{x_1^2y_2^2\over 2M^4}\ ,\\
\dot y_{2}&=&-3zy_{2}-{2x_{2}y_{2}\over x_{1}}-{\Omega x_{1}^{2}y_{1}^{3}\over \alpha M^{2}}+{3\xi M^{2}y_{1}\over \alpha}\ ,\\
\dot x_{2}&=&-3zx_{2}+{x_{2}^{2}\over x_{1}}+{y_{2}^{2}x_{1}^{3}\over 6M^{4}}-{\Omega y_{1}^{4}x_{1}^{5}\over 6\alpha M^{6}}+{\xi x_{1}^{3}y_{1}^{2}\over 2\alpha M^{2}}\ ,\\
\dot x_{1}&=&x_{2}\ ,\\
\label{eq:fpe5}
\dot y_{1}&=&y_{2}\ .
\eea

By definition, the fixed points are determined by the solutions of the algebraic system 
\bea
\dot z=\dot y_{1}=\dot y_{2}=\dot x_{1}=\dot x_{2}=0\ .
\eea
Solving the system of equations \eqref{eq:fpe1}-\eqref{eq:fpe5} we find two family of solutions as follows

\textbf{Solution $P_1$ - Unstable fixed point:} This solution is essentially given by
    \begin{equation}
        x_1 \neq 0\ ,\ x_2=0\ ,\ y_1=y_2=0\ ,\ z^2= \frac{3M^2}{4\alpha}\ .
    \end{equation}
    Written in terms of the field variables, this family of solution is characterized by
\bea\label{unstablevalues}
\phi_{\rm unst}=\dot\phi_{\rm unst}=0\,,\quad \cf_{\rm unst}={\rm arbitrary}\,,\quad \dot \cf_{\rm unst}=0\,,\quad H_{\rm unst}={\sqrt{3}M\over 2\sqrt{\alpha}}\,.
\eea
Linearising the equations of motion around the above fixed point and solving for $\phi$ yields
\bea
\phi(t)=c_{1}\,e^{p_{+}Mt}+c_{2}\,e^{p_{-}Mt}\ \quad \text{where},\quad p_{\pm}={\sqrt{3}\over 4\sqrt{\alpha}}\left(-3\pm\sqrt{16\xi+9}\right)\,
\eea
and $c_{1,2}$ are arbitrary constants. Note that, for $\alpha>0$ and $\xi>0$, the exponents of our solution $p_{+}>0$ and $p_{-}<0$. It is precisely the exponentially growing mode which renders this fixed point locally unstable.
The solution for $\cf$ is, 
\bea
\cf(t)=d_{1}+d_{2}\,e^{-{3\sqrt{3}\over 2\sqrt{\alpha}}Mt}\,,
\eea
with $d_{1,2}$ arbitrary, thus $\cf$ (and so $\tilde{\omega}$) decay rapidly around the unstable fixed point.

\textbf{Solution $P_2$ - Stable fixed point:} This attractor class of solution is described by
\begin{equation}
    x_1={\sqrt{3 \xi} M^{2}\over \sqrt{\Omega}\, y_{1}}\ ,\ x_2=0\ ,\ y_1 = \text{arbitrary}\ ,\ y_2=0\ ,\ z={\sqrt{3\lambda}M\over 2\sqrt{\Omega}}\ .
\end{equation}
In terms of the field variables, the above can be recast as
\bea\label{stablevalues}
&&\cf_{\rm st}={\sqrt{3 \xi} M^{2}\over \sqrt{\Omega}\, \phi_{\rm st}}\,,\quad \dot \cf_{\rm st}=0\,, \quad \phi_{\rm st}={\rm arbitrary}\,,\quad \dot\phi_{\rm st}=0\,,\\ \non
&& H_{\rm st}={\sqrt{\alpha\lambda}\over \sqrt{\Omega}}H_{\rm unst}={\sqrt{3\lambda}M\over 2\sqrt{\Omega}}\,.
\eea
Note that, when $\alpha=\xi^{2}/\lambda$ (as in \cite{Rinaldi:2015uvu}), one has $H_{\rm unst}=\sqrt{2}H_{\rm st}$.
The arbitrariness in $\phi_{\rm st}$ at the stable fixed point, also makes $\cf_{\rm st}$ arbitrary. However, in the Jordan frame, at the stable fixed point the value of the field $\phi$ is related to the Planck mass, the latter being generated by the spontaneous breaking of the scale invariance. Since the Planck mass is the same in the two frames (see e.g. \cite{2010LRR133D}) it is meaningful to identify, at the fixed point
\bea
{M^{2}\over 2}\equiv \frac16{\xi\phi_{\rm st}^{2}}\,,
\eea
which leads to
\bea
\cf_{\rm st}={\xi M\over \sqrt{\Omega}}\ .
\eea
The dynamics of the scalar fields $\cf$ and $\phi$ around the stable fixed point are governed by the following system of coupled second order differential equations,
\bea
\ddot\phi+{3M\sqrt{3\lambda}\over 2\sqrt{\Omega}}\dot\phi+{6M^{2}\sqrt{3}\sqrt{\Omega}\over \alpha\sqrt{\xi}}\cf-{6\sqrt{3\xi}\,M^{3}\over \alpha}&=&0\ ,\\\non\\\non
\ddot \cf+{3\sqrt{3\lambda}\,M\over 2\sqrt{\Omega}}\dot \cf+{3M^{2}\xi^{2}\over \alpha \Omega }\cf-{3M^{3}\xi^{3}\over \alpha\Omega^{3/2}}&=&0\,,
\eea
with solutions given by
\bea
\cf(t)&=&\cf_{\rm st}+c_{1}\,e^{Q_{+}t}+c_{2}\,e^{Q_{-}t}\\\non\\\non
\phi(t)&=&\sqrt{3\over \xi}M+d_{1}\,e^{Q_{+}t}+d_{2}\,e^{Q_{-}t}+d_{3}\,e^{-{3\sqrt{3\lambda}\over 2\sqrt{\Omega}}Mt}
\eea
where the exponents
\bea
\label{eq:stableexponent}
Q_{\pm}={\sqrt{3}M\over 4\sqrt{\alpha}\sqrt{\Omega}}\left(-3\sqrt{\alpha\lambda}\pm\sqrt{9\alpha\lambda-16\xi^{2}}\right)
\eea
are always negative and $c_{1,2}$, $d_{1,2,3}$ are arbitrary constants of integration. Demanding that $\phi(t)$and $\cf(t)$ are real implies that ($\Omega$ is positive definite)
\begin{equation}
\label{eq:imp-constraint}
    9\alpha \lambda \geq 16 \xi^2\ .
\end{equation}
However, we anticipate that inflationary constraints impose that $0<\alpha\lambda \lesssim 0.1547\,\xi^2$ (see Sec.\ 5). This further implies that $\cf$ and $\phi$ are real if, and only if, $c_{1,2}=0$ and $d_{1,2}=0$.

Note that in the above analysis, in either case the Hubble parameter $z$ (or equivalently $H$) is fixed by the constraint equation \eqref{aux}. 
\begin{figure}[H]
\centering 
	\includegraphics[scale=0.30]{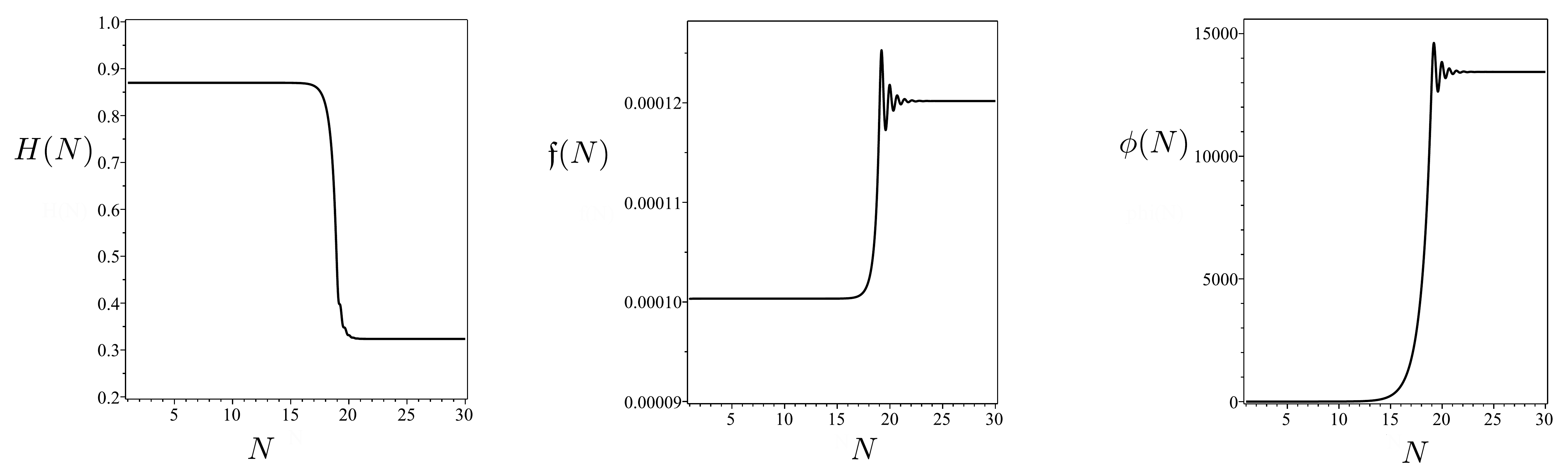}
	\caption{\it Plot of Hubble parameter $H$, scalar fields $\cf$ and $\phi(N)$ with respect to e-folding time $N$}
\label{fig:Hfphi-plot}
\end{figure} 
In Fig. \ref{fig:Hfphi-plot} we  plot  the functions $H(N), \cf(N)$ and $\phi(N)$ as functions of the e-folding time $N=\ln a$ by numerically solving \eqref{eq:FLRWmetric}-\eqref{eq:kleingordoneq} with the parameter values $\alpha=1$ and $\xi=\lambda=0.15$. These parameter values are compatible with the inflationary constraints found below. 

\section{Fields redefinition}\label{fieldsred}

To analyse inflation efficiently, we adopt the scalar field redefinition found in \cite{Tambalo:2016eqr}. The central goal of this redefinition is to obtain a potential described by a single effective scalar field. Let us define the two fields
\bea\label{defzeta}
\zeta&=&\sqrt{6} M \,{\rm arcsinh}\left(\cf\phi\over \sqrt{6}M^{2}\right)\,,\\\label{defrho}
\rho&=&{M\over 2}\ln\left({\phi^{2}\over 2M^{2}}+{3M^{2}\over \cf^{2}}\right)\ .
\eea
The Lagrangian density \eqref{eq:finalL} takes the form
\bea\label{redef_lagr}
\mathcal{L}&=& \sqrt{g}\left[{M^{2}\over 2 }R-\frac12(\partial\zeta)^{2}-3\cosh^2\left(\zeta\over \sqrt{6}M\right)(\partial\rho)^{2}-U(\zeta)\right]\,,
\eea
where the potential is now
\bea\label{poten}
U(\zeta)=-{9\xi M^{4}\over \alpha}\sinh^2\left(\zeta\over \sqrt{6}M\right)+{9\Omega M^{4}\over \alpha}\sinh^4\left(\zeta\over \sqrt{6}M\right)+{9M^{4}\over 4\alpha}\,.
\eea
As claimed earlier, the potential now is entirely governed by the effective scalar field $\zeta$ which is non-linearly related to $\cf$ and $\phi$ via \eqref{defzeta} and \eqref{defrho}. 
Such a potential has two local minima given by\footnote{One can verify that the second derivative is always positive at $\zeta_{\rm min}$ if $\lambda,\alpha,\xi$ are positive} 
\bea\label{minima}
\cosh\left(\zeta_{\rm min}\over \sqrt{6}M\right)={\sqrt{{2\Omega+\xi}\over 2\Omega}}
\eea
and a local maximum at $\zeta_{\text{max}}=0$. The values of the potential at the maximum and at the minimum are, respectively
\bea
U_{\rm max}={9M^4\over 4\alpha}\,,\quad U_{\rm min}={9\lambda M^{4}\over 4\Omega} \,.
\eea
We will shortly show that the non-zero value of the potential at the minimum is crucial for the inflationary phenomenology.

{The Friedmann equations now read 
\bea\label{newfried}
3M^2H^2&=&{\dot \zeta^2\over 2}+3\cosh\left(\zeta\over \sqrt{6}M\right)^2\dot\rho^2+U(\zeta)\,,\\\non
2M^2\dot H&=&-\dot\zeta^2-6 \cosh\left(\zeta\over \sqrt{6}M\right)^2\dot\rho^2\,,
\eea
while the Klein-Gordon ones become
\bea\label{newkg}
&&\ddot \zeta+3H\dot \zeta-{\sqrt{6}\over 2M}\sinh\left(2\zeta\over \sqrt{6}M\right)\dot\rho^2+{dU(\zeta)\over d \zeta}=0\,,\\\non
&&\ddot \rho+3H\dot\rho+{2\over \sqrt{6}M}\tanh\left(\zeta\over \sqrt{6}M\right)\dot\zeta\dot\rho=0\,
\eea 
From the above two equations it is evident that $\rho= \text{constant}$ is a trivial solution of the system, and this reinforces the fact that the inflationary dynamics is fully ruled by one field only.
This can also be seen by simply adopting the standard slow-roll approximation in the above equations. By demanding that, during slow-roll, $|\dot H|/H^2\ll 1$ one finds immediately that both the terms $\dot\zeta^2$ and $\cosh(\zeta/\sqrt{6}M)^2\dot\rho^2$ must be much smaller than $U(\zeta)$. This implies that $3M^2H^2\simeq U(\zeta)$ and that $3H\dot\zeta\simeq dU(\zeta)/d\zeta$, from which one computes the slow-roll parameters in the standard way. 

One may finally wonder if the $\rho=$ const solution is the only one with an inflationary evolution. We now show that $\rho$ = const is actually a stable attractor for the system. To see this, we first convert the system time parameter form $t$ to $N=\ln a$. Then, the second equation of \eqref{newfried} and the two equations \eqref{newkg} become, respectively
\bea\label{newdynsys}
H'&=&-H\left({A^2\over 2M^2}+{3C(\zeta)^2\sigma^2\over M^2}\right)\,,\\\non\\\non 
A'&=&-\left(3-{A^2\over 2M^2}-{3C(\zeta)^2\sigma^2\over M^2}\right)A+{\sqrt{6}\over 2M }S(\zeta)\sigma^2-{1\over H^2}{dU\over d\zeta}\,,\\\non\\\non 
\sigma'&=&-\left(3-{A^2\over 2M^2}-{3C(\zeta)^2\sigma^2\over M^2}+{2T(\zeta)A\over\sqrt{6}M}\right)\sigma\,,\\\non\\\non 
\rho'&=&\sigma\,,\\\non\\\non 
\zeta'&=&A\,,
\eea
where  $C(\zeta)=\cosh\left[\zeta/( \sqrt{6}M)\right]$, $T(\zeta)=\tanh\left[\zeta/( \sqrt{6}M)\right]$, $S(\zeta)=\sinh\left[2\zeta/( \sqrt{6}M)\right]$, and a prime denotes a derivative with respect to $N$. For the system written in this form is trivial to find the fixed point, which are given by
\bea\label{stab}
A=0\,,\quad \sigma=0\,,\quad \zeta=\sqrt{6}M\,{\rm arccosh}\left(\sqrt{2\Omega+\xi\over 2\Omega}\right)\,,\quad H={M\over 2}\sqrt{3(\Omega-\xi^2)\over \alpha\Omega}
\eea
and
\bea\label{unst} 
A=0\,,\quad \sigma=0\,,\quad \zeta=0\,,\quad H={\sqrt{3}M\over 2\sqrt{\alpha}} 
\eea 
where we have also used the first Friedmann equation \eqref{newfried} to compute the values of $H$ and $\zeta$.
Upon linearization, we see that the equation for $\sigma$ is simply $\sigma'=-3\sigma$ for both fixed points, thus $\sigma=$ 0 (that is $\rho=$ const) is a stable attractor in both cases.
Consistently with what we found in the previous section, the fixed point \eqref{stab} is a stable attractor and corresponds to \eqref{stablevalues}, as one can see by using the field redefinition \eqref{defzeta}-\eqref{defrho}\footnote{One also need the formula ${\rm arccosh}\sqrt{1+\frac{x}{2}}={\rm arcsinh}\sqrt{\frac{x}{2}} $.}. The fixed point \eqref{unst} turns out to be a saddle point, and corresponds to the point \eqref{unstablevalues}. In summary, this analysis implies that also in the $(\rho,\zeta)$ representation the system has only two fixed points characterised by $H=$ const and $\rho$ = const. Of these, however, only one is stable.
The above analysis confirms that the field $\rho$ is constant and arbitrary throughout the whole inflationary evolution. 

This can be further  checked numerically: we have also solved the system \eqref{newfried},\eqref{newkg} by choosing initial conditions nearby the unstable fixed point and we plotted the results in Fig. \ref{dynsysplots}. As one can see, the value of $\rho'$ rapidly drops to zero after just a few e-foldings. Analogously the values of $H$ and $\zeta$ follow slightly different patterns but reach the same fixed point value reported in eqs. \eqref{stab}.  We checked that the results are unchanged by choosing different values of $\rho(0)$, thus confirming that the value of $\rho(N)$ is constant, completely arbitrary and does not affect the overall dynamics.

\begin{figure}[H]
	\centering 
	\includegraphics[scale=0.26]{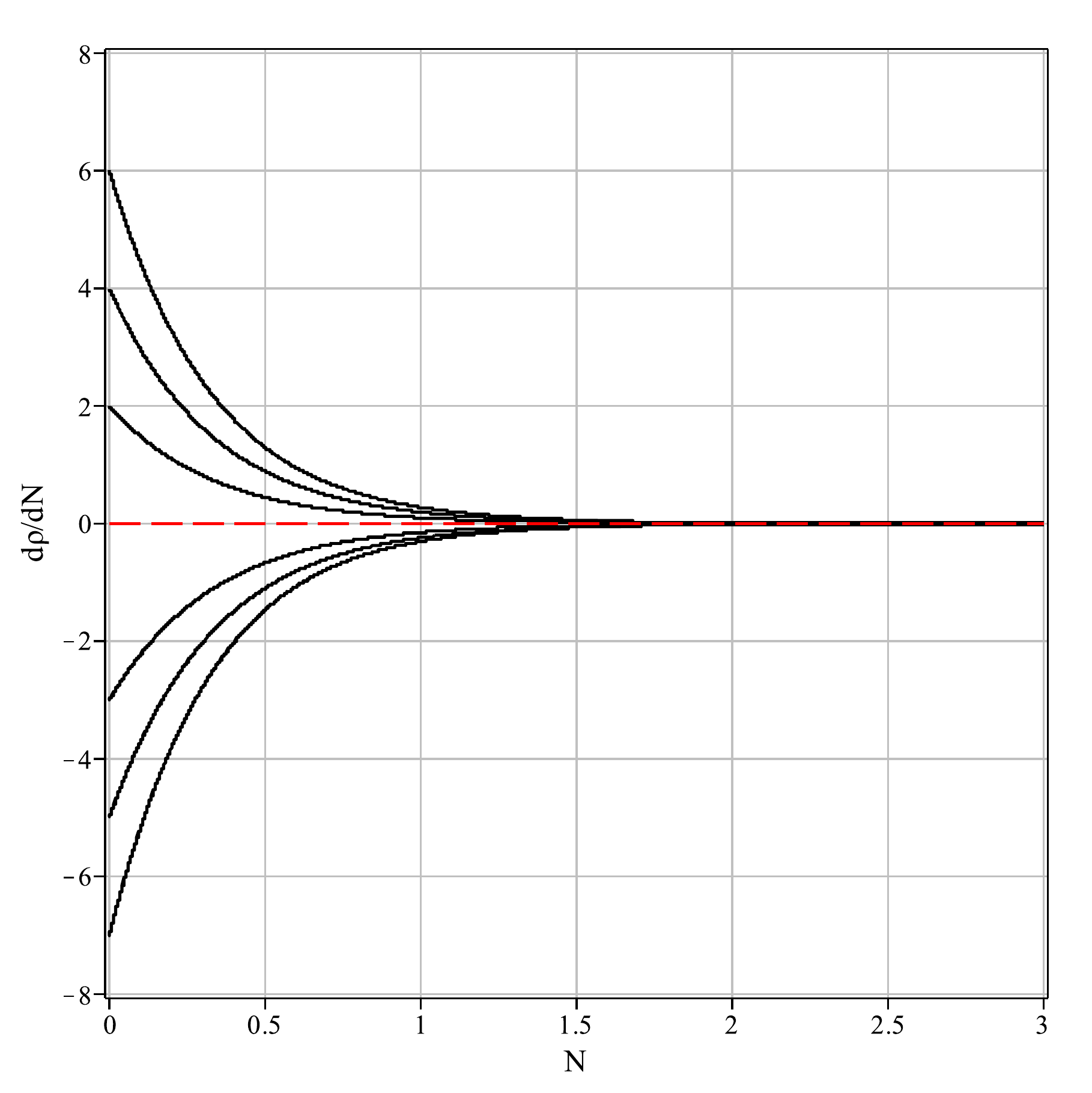} 
 \includegraphics[scale=0.26]{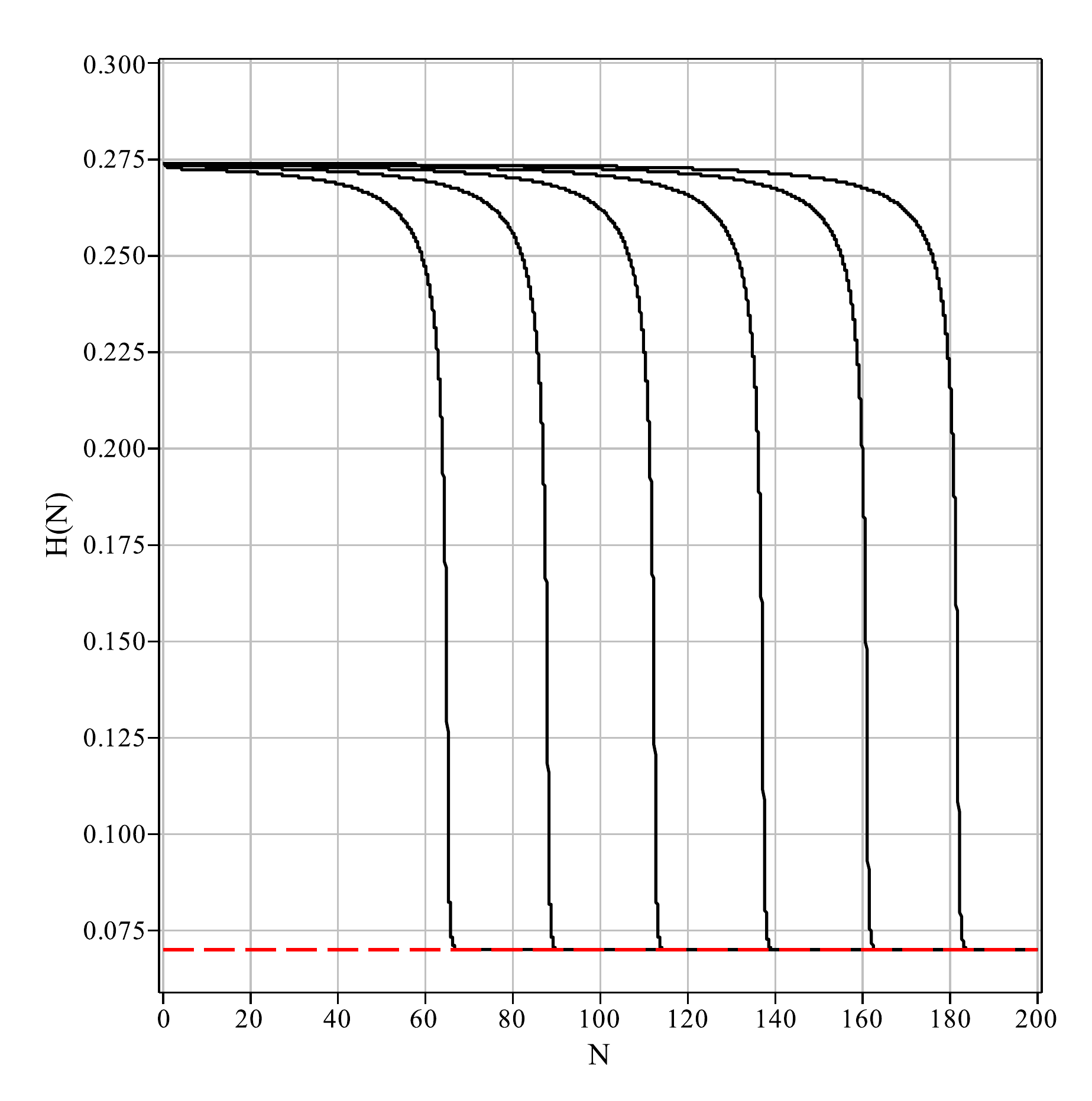}
 \includegraphics[scale=0.26]{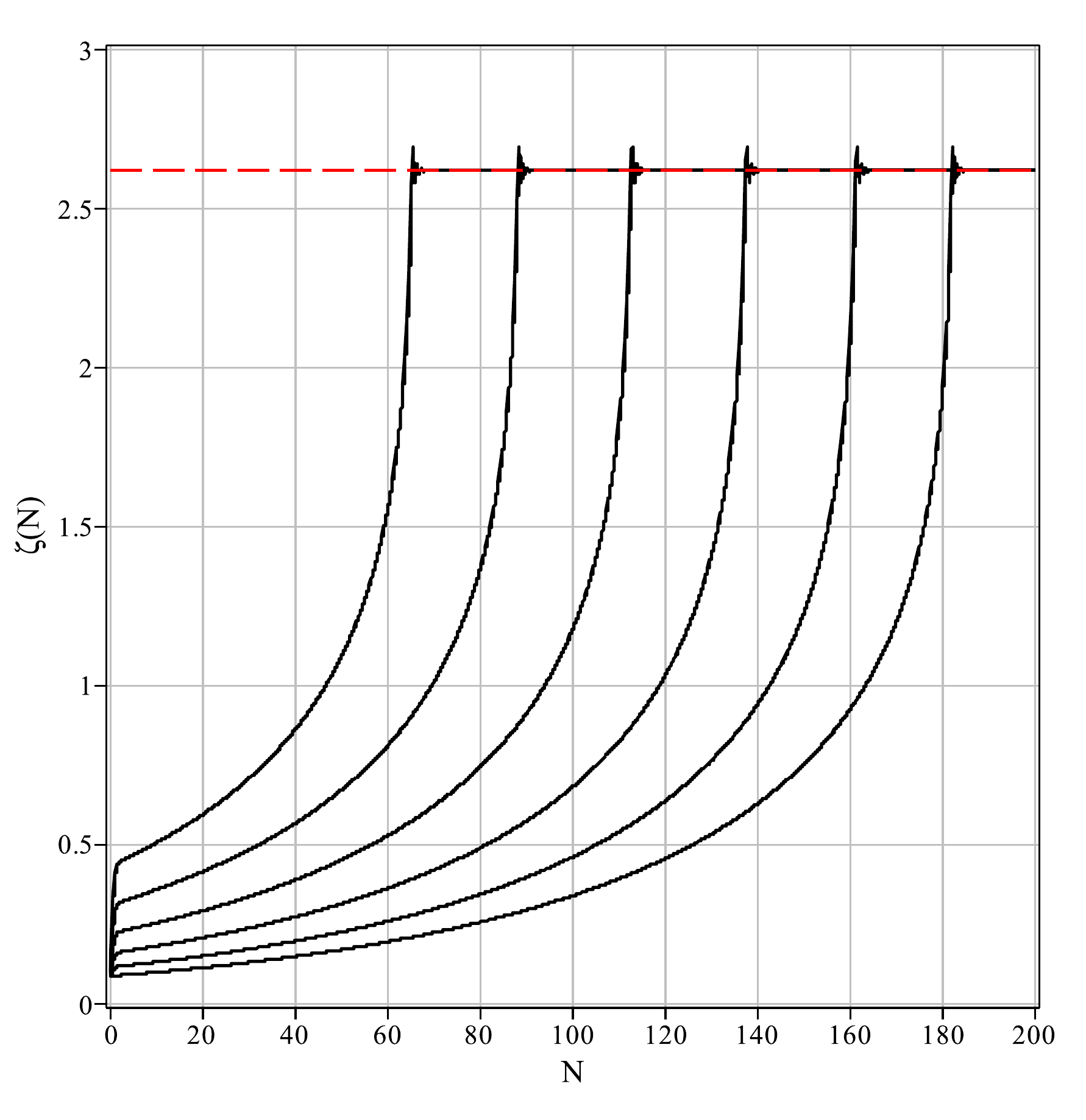}
	\caption{\it Plots of the curves $\rho'(N),H(N),\zeta(N)$ resulting from the system \eqref{newfried},\eqref{newkg}. The red dashed curves are the asymptotic values, corresponding to the stable fixed point defined by eqs.\ \eqref{stab}. The six curves have six different initial values for $\rho'(0)$ (see the first plot). As for the remaining initial conditions we chose $\rho(0)=1.47$, $\zeta(0)=0.25$, $\zeta'(0)=-0.25$, $H(0)=0.27$. All variables have been rescaled so that $M=1$. As for the parameters, we have chosen $\xi=0.01,\,\Omega= 1.07\xi^2,\, \alpha = 10$, as at the end of Sec.\ \ref{infl}.}
	\label{dynsysplots}
\end{figure}

\section{Inflation}\label{infl}

We now analyze the system to ascertain whether our model is suitable for a consistent inflationary scenario. Since an exact analytical study is not feasible, we will consider two opposite inflationary regimes, namely one that possibly occurs when $\zeta/M\ll 1$  and the other when $\zeta/M\gtrsim 1$.

First though, we study the slow-roll parameters in the inflationary potential given by \eqref{poten} as generally as possible. The first two slow-roll parameters following from \eqref{poten} are
\bea\label{e1}
\epsilon&=&{M^{2}\over 2}\left({1\over U}{dU\over d\zeta}\right)^{2}={16y(y+1)(2\Omega y-\xi)^2\over 3(4\Omega y^2-4\xi y+1)^2}\,,\\\non\\\label{e2}
\eta&=&{M^{2}\over U}{d^{2}U\over d\zeta^{2}}={4\left[8\Omega y^2+2y(3\Omega -\xi)-\xi\right]\over 3(4\Omega y^2-4\xi y+1)}\,,
\eea
where,
\bea
y=\sinh\left(\zeta\over\sqrt{6}M\right)^2\,.
\eea 
We note here that the parameter space is given by $\xi$ and $\Omega$. In the limit of large $\Omega$, both slow-roll parameters become independent of $\xi$. In fact, it is then straightforward to verify both $|\epsilon|,|\eta| >1$ for arbitrary values of $\zeta$, thus not allowing inflation to occur in this parameter regime.

Since, $\alpha$ and $\lambda$ are both positive, the condition $\Omega>\xi^{2}$ must be always implemented in the subsequent analysis. An important constraint comes from the fact that,
\bea
\lim_{\zeta\rightarrow 0}\eta=-\frac43 \xi\,,
\eea
which implies that a slow-roll kind of approximation is good only if $\xi<3/4$, as it is consistent with the slow-roll condition $|\eta|<1$. In our case, $\eta$ is negative only for small $\zeta$ but becomes positive before inflation ends. If one uses the Hubble flow parameters instead \cite{Martin:2013tda}, the constraint on $\xi$ is tighter, namely $\xi<3/8$, which we will use from now on. 
Recall that, for slow-roll inflation, the spectral indices are given by,
\bea
n_{s}\simeq1+2\eta-6\epsilon\,,\quad r=16\epsilon\,,
\eea
and we can initially estimate the range of the parameters $\Omega$ and $\xi$ in the small and large field limit.

\subsection{Small field limit}

By expanding to order $\zeta^{2}/M^{2}$ the two spectral indices $n_s$ and $r$ can be related as
\bea
r\simeq{16\xi^{2}(1-n_{s}+8\xi)\over 9\Omega-8\xi^{2}-3\xi}\,.
\eea
In Fig. \ref{omVSxi}, we have plotted $\Omega$ as a function of $\xi$ by assuming the above relation with $n_{s}=0.96$ and $r=0.03$ (recently measured upper limit, denoted by brown curve) and two smaller values of $r=0.003$ (blue curve) and $r=0.0003$ (green curve). The range of $\Omega$ increases with smaller value of $r$ within the constraint  $0<\xi<3/8$. Let us keep in mind that this results is only indicative as it refers to the case when inflation occurs for small $\zeta/M$ and, furthermore, it does not take into account yet how long inflation lasts. 

\begin{figure}[H]
	\centering 
	\includegraphics[scale=0.4]{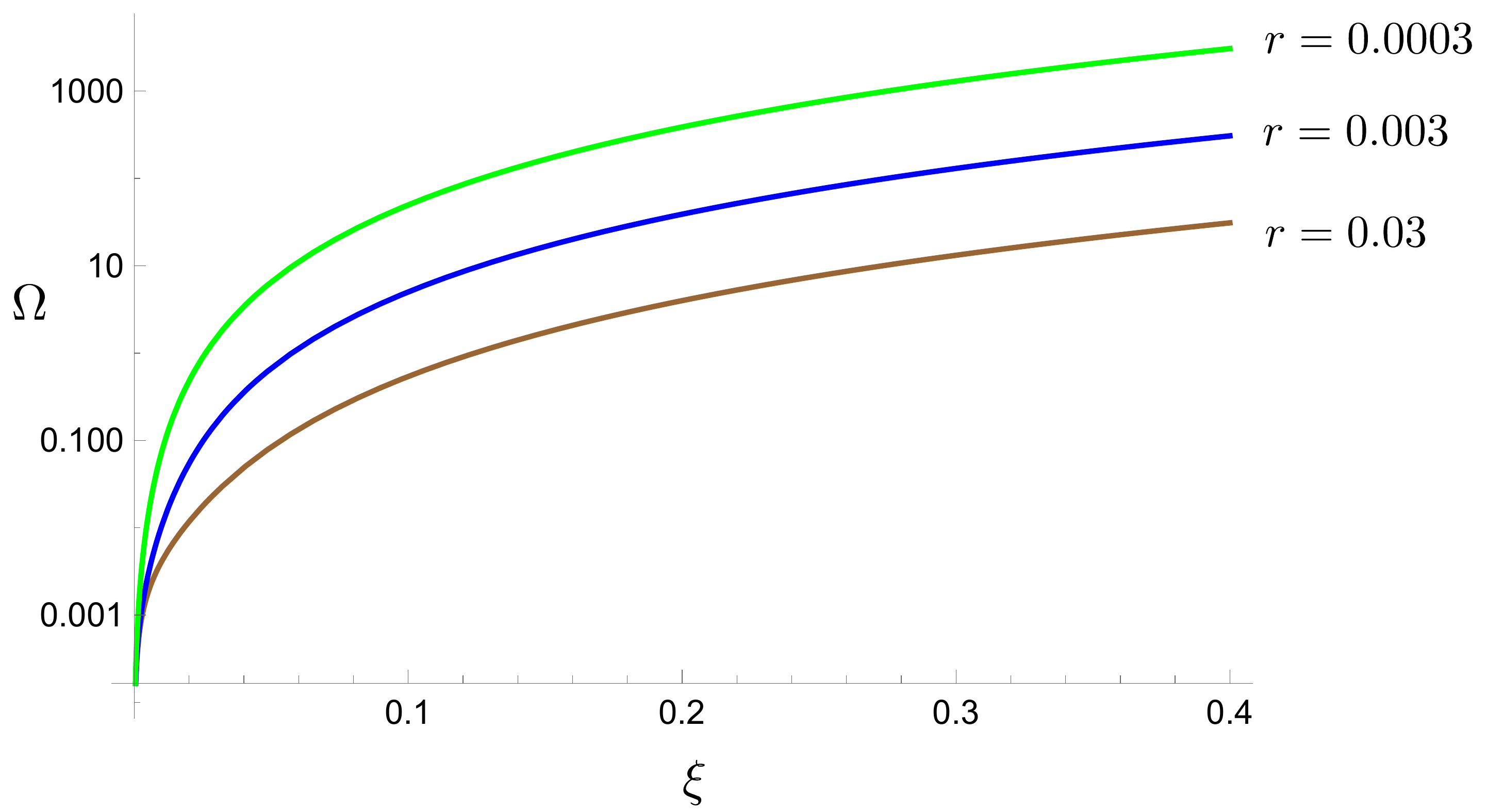} 
	\caption{\it Plot of $\Omega$ vs $\xi$ parameter space for $n_s=0.96$ and $r=0.03, 0.003, 0.0003$ corresponding to brown, blue and green curves respectively.}
	\label{omVSxi}
\end{figure}

To check the viability of this area of the parameter space we compute the duration of inflation in terms of the number of e-foldings $\Delta N$. We simply use the standard formula,  in the small $\zeta/M$ limit, to connect the values of the field $\zeta_{*}$ at the beginning and at the end of inflation $\zeta_{\rm end}$, namely
\bea\label{efold}
\Delta N\simeq-{1\over M^{2}}\int_{\zeta_{*}}^{\zeta_{\rm end}}{{U\over U'}d\zeta}\,.
\eea
The end of inflation conventionally occurs at $\epsilon=1$, which yields, in our approximation, 
\bea
\zeta_{\rm end}\simeq{3\sqrt{2}M\over 4\xi}\,.
\eea
By using \eqref{efold} with the potential expanded to order $\zeta^{2}/M^{2}$, we find that 
\bea
\zeta_{*}\simeq \zeta_{\rm end}\,\exp{\left(-{4N\xi\over 3}\right)}\,.
\eea
Note that $\zeta_{\rm end}$ is inversely proportional to $\xi$ and thus, even for the largest possible $\xi$ we find that the smallest possible $\zeta_{\rm end}=2.83\, M$, while, with $N=60$,  $\zeta_{*}\sim 10^{-13}M$. This is suggestive of a breakdown in the small $\zeta/M$ approximation during inflation rendering the above analysis unreliable.

There is an alternative argument for excluding the small field limit as a candidate for viable inflationary scenario. In the limit $\zeta\ll M$, we can expand the potential around the maximum to find
\bea
\label{eq:smallfieldV}
V \approx {9M^4\over 4\alpha}\left[1-{2\xi\over 3}{\zeta^2\over M^2}-\frac{1}{9}\left({\xi\over 3}-\Omega\right){\zeta^4\over M^4}+{\cal O}\left(\zeta^6\over M^6\right)\right]
\eea
which is essentially the so-called hilltop inflation. However, Planck results strongly favours the quartic model over the quadratic one. This means that, in our case, $\xi$ should be vanishing but then the coefficient of the quartic term would be negative, turning the local maxima into a local minima. 

Examining this case a bit closer, we first compute the slow-roll parameters $\epsilon$ and $\eta$ using the above form of the potential in \eqref{eq:smallfieldV}. We find that for $\epsilon=1$ to have real solutions, one must set $\Omega>4/3$. Further, setting $\xi=0$, we find 
\bea
n_{s}={{81+y^6(y^2-24)\Omega^2+18y^2(y^2+12)\Omega}\over (\Omega y^{4}+9)^{2}}\,,\quad r={128\Omega^2y^6\over (\Omega y^4+9)^2}\, ,
\eea
where $y=\zeta/M$. By computing $\zeta_{*}$, i.e. the value of the field at the beginning of inflation with the help of \eqref{efold}, we find that $\zeta_{*}\ll M$, in turn implying $y_{*}\ll 1$. In this limit, we have
\bea
n_{s}\simeq 1+{8\over 3}\Omega y^{2}\,,
\eea
which is clearly incompatible with data, since $\Omega>4/3$ leading to $n_s>1$.

\subsection{Large field limit}
The above discussion on the small field limit shows that we need a different kind of approximation in order to have inflation. We note that $\epsilon$ vanishes, by construction, at the local minimum of the potential $U$. In contrast, for a typical Mexican hat potential of the form $V=(\phi^2-\phi_0^2)^2$ the parameter $\epsilon$ diverges at the minimum located at $\phi=\phi_0$. Then, if inflation begins off the local maximum at $\phi=0$ (like, e.g., in hilltop models) it certainly ends at a value of $\phi_{\rm end}<\phi_0$, where $\epsilon(\phi_{\rm end})=1$.
\begin{figure}[H]
	\centering 
	\includegraphics[scale=0.50]{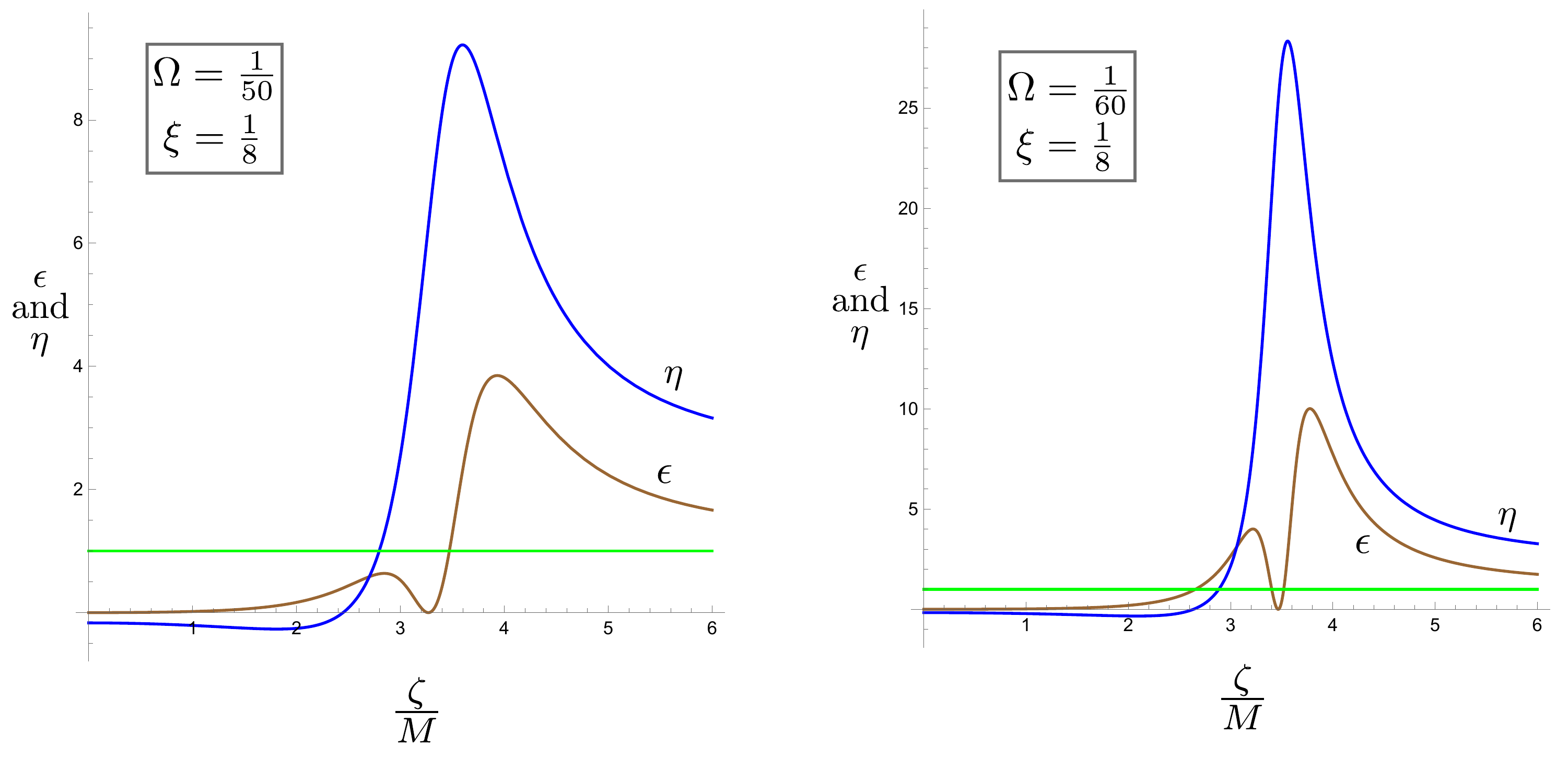} 
	\caption{\it Plot of the spectral indices $\eta$ and $\epsilon$ with respect to the dimensionless quantity $\zeta/M$ for $\Omega=1/50$ and $\xi=1/8$ on the left and $\Omega=1/60$ and $\xi=1/8$ on the right.}
	\label{etaVeps}
\end{figure} 

We note an interesting point regarding the behavior of the slow-roll parameters plotted above. For the point $\Omega=\frac{1}{64}$ and $\xi=\frac{1}{8}$, we see that $\epsilon$ and $\eta$ both diverges. This behavior can be traced back to the explicit forms of $\epsilon$ and $\eta$ as given in \eqref{e1} and \eqref{e2}. From these equations, we see that, for $\xi =\frac{1}{8}$, the denominator does not have any real valued zero for $\Omega< \frac{1}{64}$. For $\Omega=\frac{1}{64}$, $y=\sinh \frac{\zeta}{\sqrt{6}M}=4$, however, we see that the denominator of \eqref{e1} (and hence \eqref{e2}) vanishes leading to a divergence. Thus, it seems that beyond these parameter regimes, there will always exist a real value of $\zeta$ where the slow-roll parameters will blow up. This implies an upper bound on $\Omega$ consistent with the inequality imposed by \eqref{eq:imp-constraint}.

We now return to our discussion on inflation in the large field limit. The phenomenology differs substantially since the potential does not vanish at the minimum while the first slow-roll parameter does. This implies that, eventually, the inflaton field can roll over and beyond the minimum of the potential before the end of inflation. We first show that $\epsilon$ must have a local maximum between $\zeta=0$ and $\zeta=\zeta_{\rm min}$. In fact, the potential has a local minimum at $\zeta_{\rm min}$, defined by \eqref{minima}. Here, we obviously have
\bea
{dU\over d\zeta}\Big|_{\zeta_{\rm min}}=0\,,\quad {d^{2}U\over d\zeta^{2}}\Big|_{\zeta_{\rm min}}>0\quad \Rightarrow \quad \epsilon(\zeta_{\rm min})=0\quad {\rm and}\quad \eta(\zeta_{\rm min})>0\,.
\eea
On the other hand, at $\zeta=0$ (the maximum of the potential), we have
\bea
{dU\over d\zeta}\Big|_{0}=0\,,\quad {d^{2}U\over d\zeta^{2}}\Big|_{0}<0\quad \Rightarrow \quad \epsilon(0)=0\quad {\rm and}\quad \eta(0)<0\,.
\eea
Since $\epsilon$ is a semi-positive definite function, there exists (at least) one value $0<\bar \zeta<\zeta_{\rm min}$ where $\epsilon$ has a local maximum, that is 
\bea\label{locmax}
{d\epsilon\over d\zeta}\Big|_{\bar \zeta}=0\,.
\eea
This said, we notice that
\bea
{d\epsilon\over d\zeta}=-{\sqrt{2\epsilon}\over M}(\eta-2\epsilon)\,,
\eea
thus the condition \eqref{locmax} becomes 
\bea
\epsilon(\bar \zeta)=0\quad {\rm or}\quad\epsilon(\bar \zeta)=\frac12\eta(\bar \zeta)\,.
\eea
The second condition implies that the maximum value of the first slow-roll parameter $\epsilon$, in the range of interest, is at most one half of $\eta$, hence, if $\eta<1$ also $\epsilon<1$. Note that the condition $\Omega>\xi^{2}$ is crucial to guarantee that the potential is positive-definite so that the slow-roll parameters do not diverge anywhere. Fig. \ref{etaVeps} shows a typical plot of the two slow-roll parameters for two slighly different choices of $\Omega$ and the same $\xi$. We note that, in one case, $\eta=1$ for smaller values of $\zeta$ than in $\epsilon=1$ while in the other $\epsilon=1$ for smaller values of $\zeta$ than $\zeta_{\rm min}$ and in $\eta =1$. In the first case, the inflation seems to go through the zero of the potential at $\zeta_{\rm min}$, however this scenario is not physically viable. In fact, eq.\ \eqref{efold} can be written also as
\bea
\Delta N=-{1\over M}\int_{\zeta_*}^{\zeta_{\rm end}}{d\zeta\over \sqrt{2\epsilon}}\,,
\eea
which clearly shows that the integral is divergent if $\zeta_{*}<\zeta_{\rm min}<\zeta_{\rm end}$. In this case, inflation proceed for an infinite amount of time as $\zeta$ approaches asymptotically $\zeta_{\rm min}$.  A similar case occurs in the case of K\"ahler Moduli Inflation (see p.\ 60 of \cite{Martin:2013tda})

In view of all these considerations, it is evident that a viable inflationary scenario is realised only when the field slow-rolls from some point near the maximum of the potential and the condition $\epsilon\ll 1$ breaks for $\zeta<\zeta_{\rm min}$ that is for $\epsilon(\bar \zeta)>1$. In such a case, the equation $\epsilon=1$ has four real solutions that are algebraically extremely complicated and intractable. Nevertheless, we found a valid approximation for the regime of interest, which is consistent with the fact that, as proven before, inflation occurs in the large field limit. Thus, we modify the potential by replacing the hyperbolic sine with the positive exponential, namely
\bea\label{utilde}
U\rightarrow \tilde U={9M^4\over 4\alpha}\left[1-\xi\,\exp\left({\sqrt{6}\zeta\over 3M}\right)+{\Omega\over 4}\,\exp\left({2\sqrt{6}\zeta\over 3M}\right)\right]\,.
\eea
This approximation is particularly accurate in the area of the parameter space where the spectral indices are within the experimental bounds, as we will show below. The first slow-roll parameter reads
\bea
\epsilon(x)={4x^{2}(\Omega x -2\xi)^{2}\over 3(\Omega x^{2}-4\xi x +4)^{2}}\,,
\eea
where we set
\bea
x=\exp\left(\sqrt{\frac23}{\zeta\over M}\right)\gg 1\,.
\eea
The equation $\epsilon=1$ has the solutions
\bea\label{sole1}
x&=&{2(\sqrt{3}-1)\xi\over \Omega}\pm{2\sqrt{3}\sqrt{  (2\sqrt{3}-3)(2\sqrt{3}\xi^2-3\Omega)}  \over 3\Omega}\,,\\\non
x&=&-{2(\sqrt{3}+1)\xi\over \Omega}\pm{2\sqrt{3}\sqrt{  (2\sqrt{3}+3)(2\sqrt{3}\xi^2+3\Omega)}  \over 3\Omega}\,,
\eea
and they are all real (three positive and one negative) if, and only if, the new condition
\bea\label{Omegaconstr}
\Omega<{2\sqrt{3}\over 3}\xi^{2}\simeq 1.1547\, \xi^{2}\,,
\eea
is enforced. This new bound, together with the previous condition $\Omega>\xi^{2}$, substantially shrinks the parameter space $(\Omega,\xi)$ into a narrow band. In terms of the original parameter, we thus have $0<\alpha\lambda<0.1547\,\xi^{2}$ as a condition to have inflation. This result also exclude the original assumption, made in \cite{Tambalo:2016eqr, Rinaldi:2015uvu}, that $\alpha=\xi^2/\lambda$ is compatible with inflation.
In addition, the constraint $\Omega<1/64$, which prevent the slow-roll parameters to diverge, implies that $\xi\lesssim 0.11$.

In terms of $x$ the first two spectral indices read
\bea
&&n_{s}={48-5x^4\Omega^2-8\Omega\xi x^3+(16\xi^2+88\Omega)x^2-160\xi x\over 3(\Omega x^2-4 x\xi+4)^2}\,,\\\non
&&r={64x^2(\Omega x-2\xi)^{2}\over 3(\Omega x^2-4x\xi+4)^2}\,.
\eea
We invert the definition of $r$ to find $x=x(r)$ and we choose the solution (there are 4) which corresponds to the first positive zero of $\epsilon=1$. We then expand for small $r$ to find
\bea
x_{r}\simeq {\sqrt{3r}\over 4\xi}\,.
\eea
By substituting $x$ with $x_{r}$ in $n_{s}$ and expanding again for small $r$ we finally find an approximate relation between $r$ and $n_{s}$, which is the same obtained for the Starobinsky model, namely
\bea\label{nsofr}
n_{s}\simeq 1-\sqrt{r\over 3}\,,
\eea
and that does not depend on the parameters $\xi$ and $\Omega$ and it is plotted in fig. \ref{nsrdata}.

We now compute the value of the field at the beginning of inflation. We consider the standard formula
\bea
\Delta N\simeq-{1\over M^{2}}\int_{\zeta_{*}}^{\zeta_{\rm end}}{{\tilde U\over \tilde U'}d\zeta}\,.
\eea
By using the variable $x$ and the expansion for $\Omega\rightarrow \xi^2$, justified by the above considerations, we find that
\bea
-{1\over M^2}\int {\tilde U\over \tilde U'}d\zeta\simeq -\frac34 \ln x-{3\over 2\xi x}\,.
\eea
Now, let $x_{i,f}$ be the values of the field at the beginning and at the end of inflation, respectively. The value of $x_f$ is the smallest positive solution among \eqref{sole1} i.e.
\bea
x_f={2(\sqrt{3}-1)\xi\over \Omega}-{2\sqrt{3}\sqrt{  (2\sqrt{3}-3)(2\sqrt{3}\xi^2-3\Omega)}  \over 3\Omega}\,,
\eea

while $x_i$ is determined by solving the equation
\bea\label{solyi}
y_i=y_f+\frac43\Delta N+{2\over \xi}\left(e^{-y_f}-e^{-y_i}\right)\,,
\eea
where $y_{i,f}=\ln{(x_{i,f})}$ and $\Delta N=N_f-N_i>0$. The solution can be expressed in terms of the Lambert function $W_{-1}$ as
\bea
y_i=y_f
+\frac43 \Delta N+\frac23\,e^{-y_f}+W_{-1}(Q)\,,
\eea
where the suffix $-1$ indicates that we are in the "lower" branch of the function and
\bea
Q=-{2\over \xi}\,e^{-\frac43 \Delta N}\,\exp{\left(-y_f-{2\over \xi}\,e^{-y_f}\right)}\,.
\eea
The lower branch must be chosen because, for $\Delta N=0$ (i.e. $y_i=y_f$), we have $W=-\frac23 x_f$ in \eqref{solyi}, which is certainly smaller than $-1$. On the other hand, also for large $\Delta N$ and $y_f>y_i$ one has $W<-1$. By using the properties of the Lambert function and restoring the variable $x$ we finally find
\bea
x_i=-{2\over \xi}\,{ W_{-1}^{-1}\left[ -{2\over \xi x_f}\,\exp\left(  -\frac43 \Delta N -{2\over \xi x_f}\right) \right]}\,.
\eea

Let's assume that inflation lasted 60 efolding, that $\Omega=1.07\, \xi^{2}$ and $\xi=10^{-2}$. Then we find
\bea\label{numbers}
x_{i}=2.33\rightarrow \zeta_{i}=1.04\,M\,,\quad x_{f}=99.78\rightarrow \zeta_{f}=5.637\,M\,,
\eea
which makes clear that the small $\zeta$ approximation is not appropriate. In addition, we find
\bea\label{indices}
n_s=0.9679\,,\quad r=0.003\,,\quad \xi_{v}^{2}=0.00024\,.
\eea
\begin{figure}[ht]
	\centering 
	\includegraphics[scale=0.53]{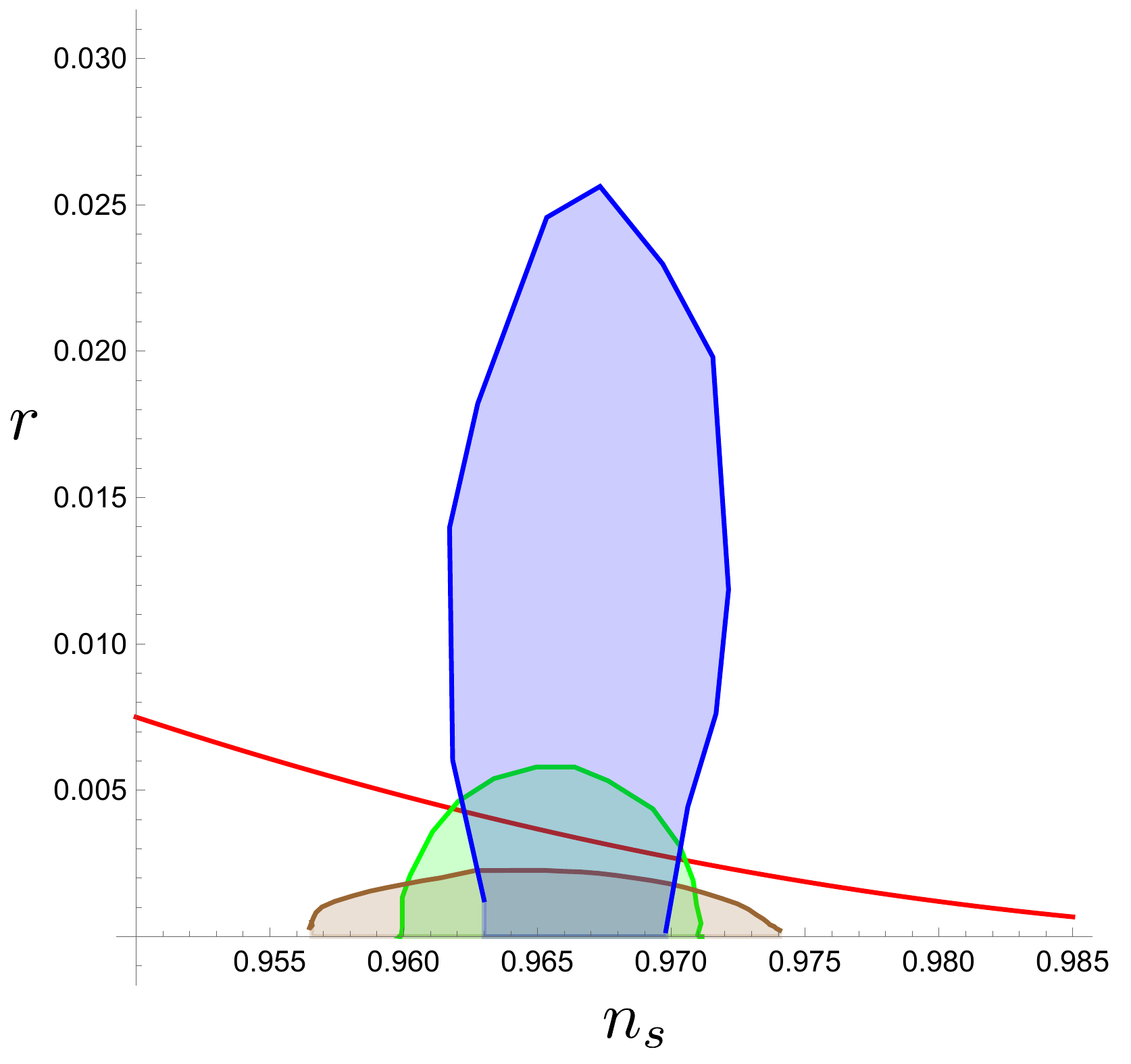} 
	\caption{\it Plot of $r$ vs $n_s$ following from \eqref{nsofr}. The blue region represents the current PLANCK + BICEP constraints \cite{Planck:2018vyg, BICEP:2021xfz, BICEPKeck:2022mhb, Campeti:2022vom}
, the green region represents the future reach of Simon's Observatory\cite{SimonsObservatory:2018koc}  while the brown region depicts the detection range of LiteBIRD \cite{LiteBIRD:2022cnt}}
	\label{nsrdata}
\end{figure}

By varying these parameters, we find that the results are mostly sensitive to the number of foldings, rather that the value of $\Omega$ and $\xi$, in line with the result \eqref{nsofr}. 

To check the consistency of our approximations, in Fig.\ \ref{test} we plot the potentials $U$ and $\tilde U$ and the corresponding first slow-roll parameters.

\begin{figure}[H]
	\centering 
	\includegraphics[scale=0.42]{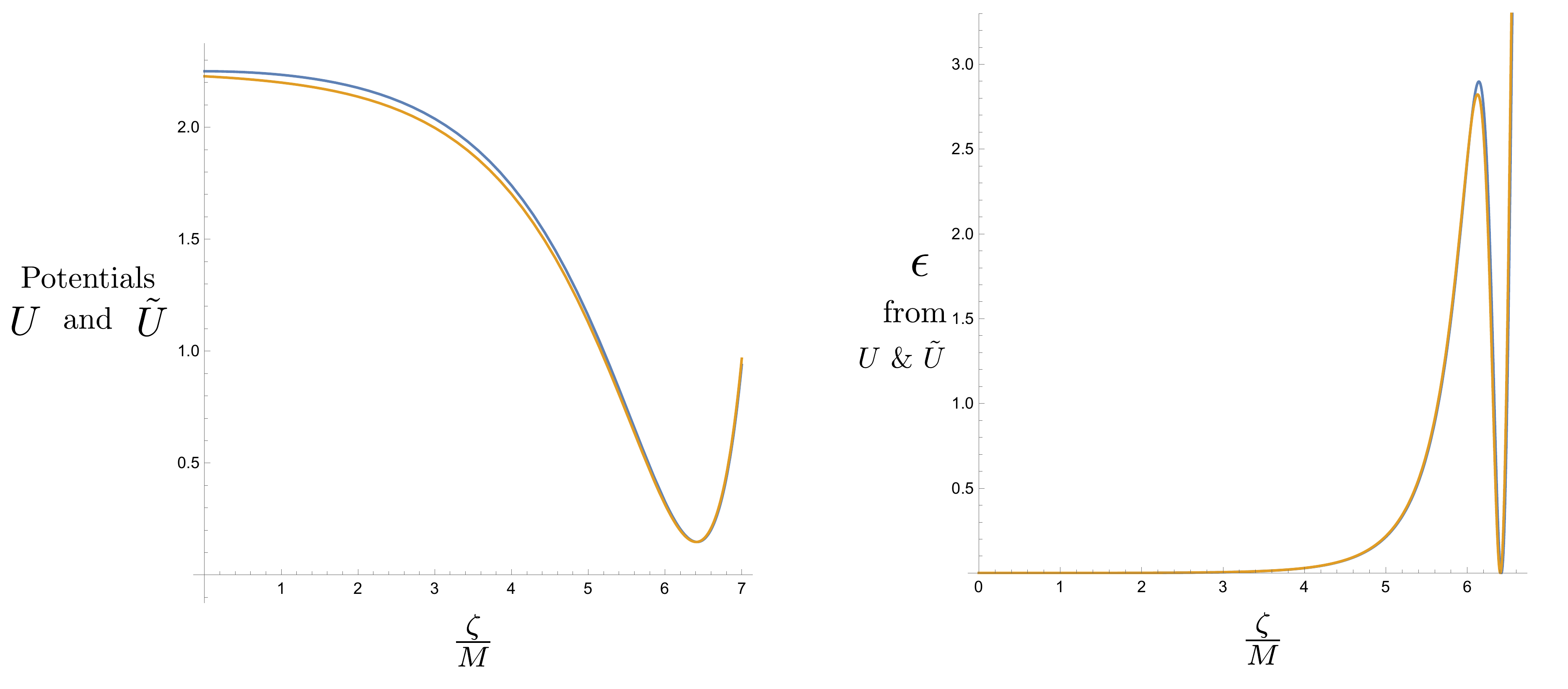} 
	\caption{\it Plot of $U$ (blue curve) and $\tilde U$ (yellow curve) on the left and of $\epsilon$ constructed with $U$ (blue curve) and $\tilde U$ (yellow curve) with  $\Omega=1.07\, \xi^{2}$,  $\xi=10^{-2}$, $\alpha=1$ and $M=1$.}
	\label{test}
\end{figure} 

\section{Tensor perturbations in scale-invariant model of gravity}\label{tpert}

Having explored inflation and the inflationary parameters in the earlier section, we now shift our focus towards  understanding tensor perturbations--in particular the tensor power spectrum in the model under consideration. It turns out to be convenient to analyse such tensor perturbations in conformal time $\eta$ which is defined as
\begin{equation}
    \eta=\int \frac{dt}{a(t)}\ .
\end{equation}
The above transforms the FLRW metric to the form
\begin{equation}
\label{eq:FRWmetric}
    ds^2=a(\eta)^2(-d\eta^2+\delta_{ij}dx^idx^j)\ .
\end{equation}
The Friedmann equations \eqref{eq:friedmanneq} becomes,
\begin{eqnarray}
\label{eq:Friedmann-1}
\cH^2&=&\frac{e^{-\sqrt{\frac{2}{3}}\frac{\tilde{\omega}}{M}}\phi'^2+\tilde{\omega}'^2}{6M^2}+\frac{3M^2}{4\alpha}a^2-\frac{a^2V}{3M^2}\ ,\\
\label{eq:Friedmann-2}
\cH^2-\cH'&=&\frac{e^{-\sqrt{\frac{2}{3}}\frac{\tilde{\omega}}{M}}\phi'^2+\tilde{\omega}'^2}{2M^2}\ .
\end{eqnarray}
where $\cH \equiv \frac{a'(\eta)}{a(\eta)}$ and `primes' denote derivative with respect to the conformal time $\eta$. The potential $V$ is given by \eqref{eq:potential}. Turning on tensor perturbations $h_{ij}$, we write down the perturbed metric as
\begin{equation}
\label{eq:pertmetric-gaugeexplicit}
    ds^2= a(\eta)^2\left[-d\eta^2+(\delta_{ij}+\bar{h}_{ij})dx^idx^j \right]\ ,
\end{equation}
and restrict ourselves to the transverse-traceless gauge given by
\begin{equation}
    \partial_i h^{ij}=h=0\ .
\end{equation}

The equations of motion for the tensor perturbations are given by,
\begin{equation}
\label{eq:final-pert-eom}
\begin{aligned}
    2M^2\partial_{\eta}^2 h_{ij}+4M^2 \frac{a'}{a}\partial_{\eta}h_{ij}-2M^2\partial^k \partial_k h_{ij}&=\left[2\left( e^{-\sqrt{\frac{2}{3}}\frac{\tilde{\omega}}{M}}\phi'^2+\tilde{\omega}'^2\right)-\frac{a^2(9M^4-4\alpha V)}{\alpha}\right.\\
    &\hspace*{4cm}\left.+4M^2\left(\frac{2a a''-a'^2}{a^2}\right)\right]h_{ij}\ .
\end{aligned}
\end{equation}
Using the two Friedman equations \eqref{eq:Friedmann-1} and \eqref{eq:Friedmann-2}, we immediately see that the RHS of the above equation vanishes identically which is a universal behaviour of inflationary scenarios whose gravity sector is governed by the standard Einstein-Hilbert action. One can of course carry out a usual momentum space analysis by decomposing $h_{ij}$ into Fourier modes as
\begin{equation}
    h_{ij}(\eta, x^i)=\frac{1}{(2\pi)^3}\int d^3 \mathbf{k}\sum_{s=+,\times}p_{ij}^s(\mathbf{k}) h^s_{\mathbf{k}}(\eta)e^{i\mathbf{k}.\mathbf{x}}
\end{equation}
where $p^s_{ij}$ are polarization tensors with $p^s_{ij}p^{s\ ij}=1$. Plugging the above mode expansion back in \eqref{eq:final-pert-eom}, we recover,
\begin{equation}
\label{eq:main-tensorpert-eqn}
    h_{\mathbf{k}}^s(\eta)''+2\cH h_{\mathbf{k}}^s(\eta)'+k^2 h_{\mathbf{k}}^s(\eta)=0\ ,
\end{equation}
where $k^2=\mathbf{k}\cdot \mathbf{k}$. Following a standard calculation \cite{Baumann:2009ds}, quantizing the fluctuations, leads to a power spectrum for the tensor perturbations given by
\begin{equation}\label{powerspectrum}
    \mathcal{P}_h= \frac{2H_{*}^2}{\pi^2 M^2}\ ,
\end{equation}
where the RHS is measured at horizon crossing.

This is somewhat expected since, although we start with higher derivative terms in the gravitational sector, we effectively map the system to an Einstein frame action with two scalars given by \eqref{eq:einsteinframe-action}. This trade-off essentially simplifies the tensor perturbations in the gravitational sector at least upto linear order. The power spectrum obtained is infact universal for inflationary scenarios whose gravitational sector is described exclusively by the Einstein-Hilbert action.

Strictly speaking the Einstein frame action \eqref{eq:einsteinframe-action} consists of two scalars $\phi$ and $\tilde{\omega}$ which might motivate someone to perform a multi-field inflationary analysis. However, as argued earlier, a field redefinition renders the potential dependent on only one of the field while the other direction $\rho$ can be treated as a \emph{flat} direction and does not affect inflationary dynamics. The analysis of tensor perturbations indeed show that the Friedman equations \eqref{eq:Friedmann-1}, \eqref{eq:Friedmann-2} conspire in a way to give us the equation of motion of $h_{ij}$ which matches with that of a single field inflationary model with Einstein-Hilbert gravity thus vindicating our analysis based on single-field inflationary models.

\medskip
\section{Discussion and Conclusion}\label{concl}

In this work we have analysed the most general scale-invariant theory with a scalar field written in the Einstein frame, and investigated the parameter space which leads to satisfactory inflationary observations. We summarize below the main results we have found.

\begin{itemize}
\item We have performed a detailed analysis of the dynamics of the scale-invariant theory \eqref{eq:fullquadraticaction} in the Einstein frame \eqref{eq:einsteinframe-action} by means of dynamical system methods. We have found that the system evolves from an unstable de Sitter space to another stable asymptotic de Sitter space point, where an effective mass scale emerges, which can be identified with the Planck mass. The transition between the two points potentially yields an inflationary expansion of the Universe.

 \item In the Einstein frame, dynamics seems to be ruled by two scalar fields, however by means of a suitable field redefinition (see \eqref{defzeta},\eqref{defrho}) we show that the action can be written in such a way that the potential depends on one scalar field only, as in \eqref{redef_lagr}. Thus the inflationary era can be studied as a single-field inflation model.
 
 \item We found a viable model of inflation when the scalar field is of order of the mass scale $M$, which is identified with the Planck mass. In this limit, we could compute the spectral indices and their running, which shows excellent agreement with observations, see \eqref{redef_lagr}. In addition, the three free parameters of the theory are significantly constrained by data but not fine-tuned, see the relation \eqref{Omegaconstr}. In particular, the non-minimal coupling parameter $\xi$ must be small, for viable inflation to occur, as opposed to the standard Higgs inflation model.
 
 \item We have computed the power spectrum of the primordial tensor perturbations \eqref{powerspectrum}  {and found, as expected, that it does not differ from the usual one in the Einstein frame.}

\item These general results have been applied to a specific concrete model where the  scalar sector features the filed $\phi$ (the scalar field whose fixed-point value generates the Planck scale) in addition to, of course,  the Starobinsky scalar $\zeta$ due to the $R^2$ term. Interestingly, however, its power spectrum is rather close to the observational bounds for $s$-inflation, giving us constraints within the ($n_s-r$) range to be within $n_s= 0.9618, r=0.0044$ from the latest PLANCK-BICEP data.
$n_s=0.9706, r=0.0026$ range (see Fig. \ref{nsrdata}). Furthermore this also leads to possibility to test this model with future observations with CMB from CMB-S4, Simon's Observatory, LiteBIRD and CMB-Bharat\cite{Adak:2021lbu}.

\end{itemize}

It is interesting to compare our work with the results presented in \cite{He:2018gyf}. Here, the action contains an additional linear term in $R$, compared to our action \eqref{action}. One could redefine the scalar field in \cite{He:2018gyf} in order to have only one  term of type $\phi^2 R$ but then the quartic potential would be different and the scale invariance would be lost. The two models also differ as in \cite{He:2018gyf} there appears to be no field redefinition able to make the potential dependent on one field only, as in our case. Finally, the predictions in \cite{He:2018gyf} are similar to ours except that we do not have a branch where the coupling parameter $\xi$ is large.

The present work has several possible future applications. For example, it would be interesting to apply the general formul$\ae$ derived here to scale-invariant models where scale-invariance is broken by quantum effects or via non-perturbative effects \cite{Donoghue:2017vvl,Liu:2018hno,Laporte:2021kyp} and in general to many scale-invariant models, other than the one considered in this paper.

\section*{Acknowledgement}

We would like to thank C.\ Cecchini, A. Salvio for numerous discussions over the course of this work. DM would like to thank Aparajita Sen for illuminating discussions regarding LiteBIRD, Simons Obervatory and CMB-Bharat observational data. AG thanks P. Michalak for reading manuscripts and comments. DM also acknowledges the hospitality of IISER Bhopal while this work was undergoing. DM's work is supported in part by the SERB core research grant CRG/2018/002373.

\bibliographystyle{JHEP}
\bibliography{References}

\end{document}